\documentclass[runningheads]{llncs}
\usepackage{graphicx}
\usepackage{caption}
\usepackage{subcaption}
\usepackage{mathtools}
\usepackage{MnSymbol}
\usepackage{hyperref}
\usepackage{etoolbox}
\usepackage{xcolor}
\usepackage{xspace} 
    \usepackage[noend]{algorithm2e}
    \SetKwIF{If}{ElseIf}{Else}{if}{:}{else if}{else}{end if}%
    \SetKwFor{While}{while}{:}{end while}%
    \SetKwRepeat{Do}{do}{while}

\newcommand{\DefineTerms}[2]{%
	\csdef{#1}{#2\xspace}%
	\csdef{#1s}{#2$\mathtt{s}$\xspace}%
}
\SetKwComment{Comment}{$\#$\ }{}
\SetKwRepeat{Do}{do}{while}

\begin{document}
    \newcommand{\ayelet}[1]{{\color{blue}\textbf{Ayelet}: #1}}
	\newcommand{\avivz}[1]{{\color{red}\textbf{Aviv}: #1}}
	\newcommand{\change}[1]{{#1}}
	\newcommand{\boldheader}[1]{\vskip 5pt \noindent{\bf\color{gray} #1}}
	\newtheorem{experiment}{Experiment}
	
	\DefineTerms{htlc}{$\mathtt{HTLC}$}
	\DefineTerms{cltv}{$\mathtt{CLTV}$}
	\DefineTerms{cltvdelta}{$\mathtt{cltv\_expiry\_delta}$}
	\DefineTerms{finalcltv}{$\mathtt{min\_final\_cltv\_expiry}$}
	\DefineTerms{htlcmsat}{$\mathtt{htlc\_minimum\_msat}$}
	\DefineTerms{maxlock}{$\mathtt{locktime\_max}$}
	\DefineTerms{lnd}{LND}
	\DefineTerms{clightning}{C-Lightning}
	\DefineTerms{eclair}{Eclair}
	\DefineTerms{maxhtlc}{$\mathtt{max\_concurrent\_htlcs}$}
	\DefineTerms{feebase}{$\mathtt{fee\_base\_msat}$}
	\DefineTerms{feeproportional}{$\mathtt{fee\_proportional\_millionths}$}
	\DefineTerms{dust}{$\mathtt{dust\_limit\_satoshis}$}

\title{Congestion Attacks in Payment Channel Networks}
%
\authorrunning{A. Mizrahi and A. Zohar}
\author{Ayelet Mizrahi \and Aviv Zohar}
\institute{The Hebrew University of Jerusalem\\ \email{\{ayelem02,avivz\}@cs.huji.ac.il}\\}

\maketitle              
\begin{abstract}
Payment channel networks provide a fast and scalable solution to relay funds, acting as a second layer to slower and less scalable blockchain protocols. 
		In this paper, we present an accessible, low-cost attack in which the attacker paralyzes multiple payment network channels for several days.
		The attack is based on overloading channels with requests that are kept unresolved until their expiration time. Reaching the maximum allowed unresolved requests (\htlcs) locks the channel for new payments.
		The attack is in fact inherent to the way off-chain networks are constructed, since limits on the number of unresolved payments are derived from limits on the blockchain. 
		We consider three versions of the attack: one in which the attacker attempts to block as many high liquidity channels as possible, one in which it disconnects as many pairs of nodes as it can, and one in which it tries to isolate individual nodes from the network. We evaluate the costs of these attacks on Bitcoin's Lightning Network and compare how changes in the network have affected the cost of attack. Specifically, we consider how recent changes to default parameters in each of the main Lightning implementations contribute to the attacks. Finally, we suggest mitigation techniques that make these attacks much harder to carry out.

\end{abstract}

\section{Introduction}

Payment channel networks such as the Lightning Network~\cite{poon2016bitcoin} are a second layer off-chain solution to the scalability problems of blockchains. 
They require participants to lock funds into channels, which then allows them to send payments to others over several hops. Altogether, they allow both a higher number of transactions as well as faster transaction resolution compared to the underlying blockchain. 

In this paper we describe and evaluate a novel attack that locks funds in channels between honest participants that are potentially far away from the attacker, giving the attacker the ability to disrupt the transfer of payments throughout the network. In contrast to previously known attacks that locked liquidity in channels~\cite{rohrer2019discharged}, the method we present here requires \emph{lower} costs as it requires the attacker to lock a smaller amount of liquidity. We evaluate these costs in the Lightning Network, where we show that spending less than half a bitcoin, the attacker can indefinitely lock up channels holding the majority of the funds currently assigned to the network. 
\change{To summarize, our contributions are as follows:
\begin{itemize}
    \item We leverage a known limitation of payment channels (their max \htlc limit) to form two different attacks on the Lightning Network (a network wide DoS attack and a more localized node isolation attack).
        \item We provide statistics on several aspects of the Lightning Network that are relevant to the attack which may be of independent interest.
        \item We provide a thorough evaluation of the amount of resources needed to conduct each attack and of the level of harm an attacker can cause.
    \item We report on small proof of concept experiments using actual lightning nodes (on a local test network) that confirm the feasibility of the attack.
    \item We propose and discuss several short-term mitigation approaches that make the attack more difficult to carry out and reduce its efficiency. 
\end{itemize}
}

Our attack is based on the inner workings of the main mechanism that makes payment channel networks possible: Hashed Time-Locked Contracts (\htlc). Essentially, as payments are set up to move along some path in the network, all channels along the path reserve some funds for the transfer that is about to take place. The number of simultaneously reserved and unresolved payments per path is limited. Our attack thus simply opens many small payment requests along extremely long paths and keeps them unresolved for as long as possible. In this way, all channels along the path are unable to relay other transfers. 

The vulnerability can be attributed to three fundamental properties of off-chain payment networks.
    
\vspace{3pt}\noindent{\bf 1. Payments are executed in a trustless manner.} Payments are executed using conditional payment contracts (in the form of transactions with \htlcs) that are exchanged between parties and are only sent to the blockchain if disputes arise. These contracts grow in size as more conditional payments are pending, and so the total number of pending payments is limited by transactions sizes that can be placed on the blockchain. Bitcoin's Lightning Network is limited to at most 483 concurrent \htlcs ~\cite{BOLT}, while Raiden~\cite{Raiden}, Ethereum's network, is limited to at most 160 due to gas costs~\cite{raiden_limit}. 

\vspace{3pt}\noindent{\bf 2. Expiration times are long.} To allow nodes to recover their funds if a malicious partner closes a channel that is part of a pending payment, \htlc expiration times have been set to allow nodes sufficient time to appeal such closures. 
In Bitcoin's Lightning Network things are even more severe: due to lower expressiveness of its scripting language, \htlc expiration times accumulate over the length of the path, reaching up to 2016 blocks -- which typically take the Bitcoin network two weeks to produce. 

\vspace{3pt}\noindent{\bf 3. The privacy of payments.} Payment Channel Networks utilize onion routing that does not allow intermediate nodes on the path to recognize where payments originate and where they are going, allowing the attacker to act with impunity. Payment privacy essentially prevents us from attributing blame to potential attackers and add mechanisms that effectively detect the attack. 

\vspace{3pt}\noindent{\bf A description of the attack.} In order to paralyze channels, 
the attacker first adds a new node to the payment network. It then identifies a route suitable to attack, considering some restrictions on the path (maximum route length, locktime of intermediate nodes, remaining \htlc capacity) and maximizing the attack benefit (to lock channels with a large amount of funds or high betweenness value).
It opens channels with the source and target of the route, and requests many tiny payments through this path, exhausting the number of simultaneously open \htlcs. Since the attacker is both the source and destination of this payment, it can choose to delay the final execution of the payment which would remove all pending \htlcs from the path. The path is then locked for long periods of time (up to several days). Just before expiration, the attacker sends an \emph{update\_failure} message to the previous node, which cancels the payment and reverts the state, avoiding a forced closure of the attacker's channel. This allows the attacker to re-run the attack once again and lock the same path for an additional period of time. 

\change{
To successfully carry out the attack, the attacker needs liquidity on its outgoing channel as well as liquidity on its incoming channel. While it is easy to open a channel and invest liquidity, this liquidity is on the attacker's side of the channel initially which is suitable for the outgoing channel. The incoming channel's liquidity needs to allow for payments toward the attacker. Liquidity can be shifted in that direction by sending an outgoing payment from the attacker, e.g., to deposit funds in some exchange or purchase goods.
}

We stress that simple mitigation attempts such as increasing the number of \htlcs allowed per channel are not very effective. First, more allowed \htlcs will imply larger settlement transactions in Lightning (larger than the current block size). Second, the attacker can easily create enough payment requests to lock many more \htlcs (each message locks an \htlc for days and requires little effort).

In this paper we evaluate the attack specifically on the Lightning Network, which is the most prominently used payment channel network. We evaluate three main attack scenarios:
First, we consider an attack on the entire network, which attempts to lock as many channels as possible and focuses on channels holding most of the funds in the network. This sort of disruption would severely hinder the volume of payments that can be sent on the Lightning Network.
The second attack scenario we consider is one that disconnects as many pairs of nodes as possible and breaks the network into separate components.
The main complexity in carrying out these attacks is picking routes in a way that respects limits on the maximal delay incurred along the path, and still targets the channels with the highest connectivity and liquidity.
Finally, the third attack variant that we evaluate targets single nodes and paralyzes all channels that connect them to the network. 

As far as we are aware, while exhaustion of the \htlcs of a channel is known to paralyze the channel, the attack that we describe has never been evaluated for its effects on the network, or on individual nodes. In particular, there are no available estimates of the cost to attackers from executing either version of the attacks we propose.\footnote{We were able to find public record describing the basic idea of the attack, on a single channel~\cite{EmelyanenkoK,Rusty}. We note that no full translation of this vulnerability to the entire network was previously considered. Due to the public nature of these posts, we did not perform a disclosure of the vulnerability to the devs.}

\change{Due to ethical concerns, we did not attack the live network. Instead, we worked with two main complementary approaches.
In the first, we use the code of actual lightning node implementations to set small local networks to test the basic mechanics underlying the attacks. We validate the behavior of nodes in a series of experiments reported in Appendix~\ref{proof}.
In the second, we perform the attacks on a simulation of the actual network based on topology data we extracted from a live node.
We provide the full description and evaluation of the attacks in its different modes in Sections~\ref{attack1}~-~\ref{attack2}.
In Section~\ref{related} we discuss prior work, including other DoS attacks on the Lightning Network.
Our attack differs in that it requires fewer resources, repeating it indefinitely does not waste fees, and it does not require a direct connection to the victim node.}


\section{Background on the Lightning Network}\label{Background}
\change{The Lightning Network is the most widely used payment channel network to date. As of October 2020, it has more than 14k nodes and 37k channels and holds a total capacity of around 1100 BTC.}
We introduce some of the basic properties of the Lightning Network.



\paragraph{Hashed Time Locked Contracts (\htlcs)} - conditional payments
which promise an intermediate node on the channel that it can receive funds if it submits a cryptographic proof (pre-image of a hash) within a given timeframe (specified as a specific chain height). Each transaction that occurs in the Lightning Network is first set up by adding an additional \htlc output to every channel on its path. Once these are set up, the payment is executed by propagating the pre-image from the payment's recipient back along the path towards the sender. Once the pre-image arrives at some intermediate node, it can essentially guarantee that it can receive the funds (if it posts the transaction with the pre-image to the blockchain). The conditional payment is then removed from the channel and is replaced by a non-conditional reallocation of the funds. 

The main problem with the approach above is that if several payments are being set up, the number of \htlcs on a channel grows. This implies that the transaction that will eventually be posted to the blockchain will be large -- setting a natural limit on the number of \htlcs that can be simultaneously open on a channel. 

\paragraph{HTLC Timeouts} - Usually, channels are set up quickly and do not wait long for the pre-image to propagate. An \emph{update\_failure} message may sometime be returned instead of the pre-image if one of the intermediate nodes cannot or will not relay the payment. 
However, malicious nodes may withhold the pre-image and not propagate it back (or alternatively not complete the channel set up with \htlcs). In such cases, \htlcs are designed to expire. This is done using a CheckLockTimeVerify (\cltv) instruction, which essentially does not allow the \htlc to be redeemed after a certain block height. In order to ensure that intermediate nodes do not lose funds, outgoing \htlcs must expire before incoming \htlcs do. Each node specifies a parameter \cltvdelta, which specifies the difference in timeouts it is willing to tolerate. 
The timeout of payments is therefore the accumulation of the \cltvdeltas from the end of the route towards its beginning (the last node's timeout is limited by a parameter named \finalcltv instead of \cltvdelta). As \cltvdeltas are typically either 40 blocks or 144 blocks, the timeouts of \htlcs can accumulate and often take days. Nodes impose a limit on the maximal timeout \maxlock, which is set to 2016 blocks (equivalent to 2 weeks). This high timeout makes the attack extremely potent. 



\paragraph{Privacy} - One of the goals of the Lightning protocol is to preserve the privacy of users -- a fact that eventually aids our attack. For example, routing payments is done via Onion Routing which helps disguise the attacker. Additionally, the expiration of \htlcs is also conveyed along payment paths and to preserve privacy, senders are allowed to add arbitrary values to the initial delay. We exploit this fact to add to the expiration delay of \htlcs (up to the allowed maximum of two weeks). 

\section{Lightning Network Analysis}\label{Analysis}
We begin our exploration of the current state of the Lightning Network by listing the default values for various parameters in the main implementations of the Lightning protocol. These are of interest since, as we show later below, most nodes use the defaults, and thus these heavily influence the state of the Lightning Network and its vulnerability to our attack.

\subsection{Default Parameter Values} \label{defaults}
The BOLT (Basis of Lightning Technology)~\cite{BOLT} specifications detail the protocol of Lightning Networks. In our work, we focus on the main three implementations: 
\lnd~\cite{lnd}, \clightning~\cite{clightning}, and \eclair~\cite{eclair}

Each of the implementations uses slightly different default values for parameters of interest. These are depicted in the table below, along with ranges or values specified in the BOLT.\footnote{We give the defaults used in mainnet. Testnet behavior differs slightly.}

\begin{center}
\setlength{\tabcolsep}{10pt}
	\resizebox{10,5cm}{!}{%
		\begin{tabular}{ c  c  c  c  c }
			\hline
			& \textbf{\lnd} & \textbf{\clightning} & \textbf{\eclair} & \textbf{BOLT} \\ \hline
			\textbf{\cltvdelta}       & 40   & 14   & 144  & -                  \\  
			\textbf{\finalcltv}       & 40   & 10   & 9    & 9
			\\	
			\textbf{\maxlock}         & 2016 & 2016 & 2016 & \textless{$5\cdot 10^8$} \\
			\textbf{\maxhtlc}         & 483  & 30   & 30   & $\le{483}$ \\
				\textbf{\dust}        & 573    & 546   & 546  & - 
			\\
			\textbf{\htlcmsat}        & 1000 & 1000 & 1    & - 
			\\ 
			\textbf{\feebase}         & 1000 & 1000 & 1000 & - 
			\\
			\textbf{\feeproportional} & 1    & 10   & 100  & - 
			\\ 
		 \hline
		\end{tabular}
	}
\end{center}

Recent changes to the defaults have in fact made our attack easier to carry out: \lnd changed their \cltvdelta default from 144 to 40 blocks (on Mar 12th, 2019)~\cite{lnd_delta_change}, which allows chaining more nodes in each path without reaching the \maxlock limit. Nodes running an old version may still hold the 144 default that was used prior to that.

Additionally, a \maxlock of 2016 was agreed upon by Lightning developers, in the 2018 Adelaide meeting to set the BOLT 1.1 specs~\cite{Adelaide}. This is an increase of previous values used in some implementations. Again, this allows for longer routes and longer expiration delays that make the attack more damaging and easier to carry out. 

\subsection{Network Statistics}\label{Statistics}
We introduce some statistics on the parameters announced by nodes in channels on the Lightning Network.\footnote{We ignore disabled channels and channels with nodes that do not reveal their policies.} In order to perform the calculations, we took snapshots of the Lightning Network mainnet. The information was obtained from a continuously running \lnd node. Our results correspond to a network snapshot taken on September 21st, 2020. We include additional analysis with snapshots taken over a period of 18 months for comparison. 

\begin{figure}[ht]
	\centering
	\includegraphics[scale=0.38, trim={0cm 10cm 1cm 0cm},clip]{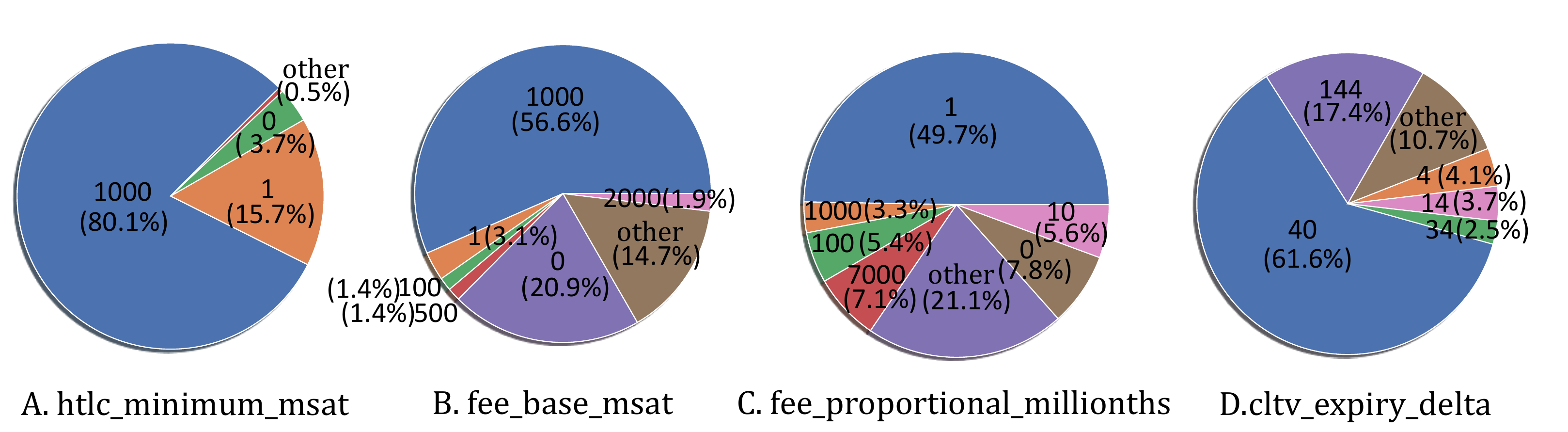}
	\caption{Statistics on parameters announced by nodes
in channels on the Lightning Network}
	\label{figure:2}
\end{figure}

In Figure~\ref{figure:2}, we present the most common values of four of the parameters announced by nodes.
It is clear that very few values are used. The remaining values appeared less than 3\% each (which we grouped together as ``other'').

Figures~\ref{figure:2}A,\ref{figure:2}B,\ref{figure:2}C show the distribution of \htlcmsat, \feebase and \feeproportional.
These represent the minimal amounts nodes are willing to transfer, the flat fee for each transfer, and the fee that grows with the transferred amount. 
These values are small relative to the default configured dust limit, which sets a threshold below which \htlcs would not be added by nodes. Therefore, these parameters have a lower impact on the cost of the attack. 
We elaborate more on costs in Section~\ref{attack1}.

Finally, we examine the distribution of \cltvdelta~- the minimum difference in \htlc timeouts the forwarding node will accept. We recall from the table in Section~\ref{defaults} that 144, 40, and 14 are the defaults that correspond to the different implementations mentioned previously. In Figure~\ref{figure:2}D, we see that the defaults constitute 82.7\% of the total.

\begin{figure}[ht]
\centering
	\includegraphics[scale=0.4]{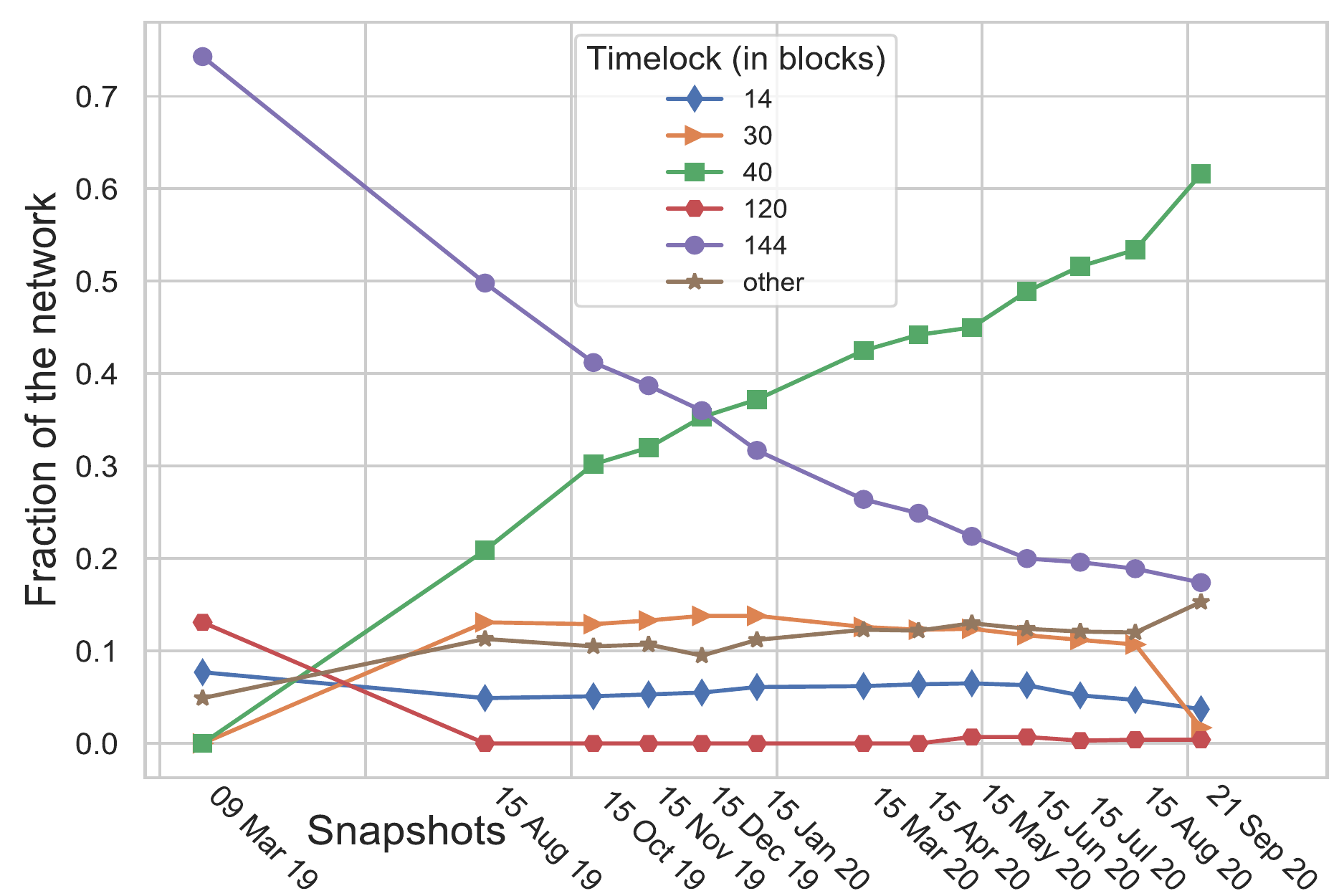}
	\caption{\cltvdeltas in different snapshots}
	\label{figure:3}
\end{figure}

\paragraph{How do values change over time?}
In our attack, the route length we can compose is often limited by the values of \cltvdelta. Figure~\ref{figure:3} shows the changes in \cltvdelta over an 18 month period\footnote{The snapshot from Mar 9th, 2019 was taken from~\cite{rohrer2019discharged,snapshots}.}. We show only the most common values. \change{The choice of presented dates was according to available information from our node and was slightly affected by downtime. Since channels are open for a long period of time, the exact day chosen does not impact the topology.}  

The main change over this period is the decreased use of the value 144 for expiry time and the increase in the use of the value 40. We attribute this to the fact that \lnd changed their default \cltvdelta from 144 to 40 in Mar 2019~\cite{lnd_delta_change}. 
In Appendix~\ref{tag}, we show that \lnd nodes are both the most common nodes (we estimate that they constitute 91.3\% of the network), and also the ones that hold most of the liquidity in the network.



\section{Attacking the Entire Network}\label{attack1}
In this section, we consider a malicious node that wishes to disrupt the entire network's operation.
\change{Initially, it connects to other nodes in the network by opening channels with them, allowing it to learn the topology of the network and launch the attack.
Then, the attacker} uses a greedy algorithm in order to pick routes and paralyze as much liquidity or to disconnect as many pairs of nodes as possible. For each route, the attacker will initiate \maxhtlc payments, and withhold the response, turning all channels along the path unavailable for new requests. Just before expiration, the attacker will announce a failure to complete the payment. This step is repeated for multiple disjoint routes making the network less and less  connected.

\change{Examining several heuristics and path selection choices we found that a greedy choice of routes which we fully paralyze one by one makes the most out of every attacker’s outgoing channel and achieves approximately optimal results for the attacker as we will present in the following evaluation.} The main challenge faced by the attacker \change{in this heuristic} is to use routes composed of channels with similar \maxhtlc so that we do not leave parts of the path unlocked, and to fit as many high-liquidity channels within the limits of 20 hops and \maxlock total delay. Hence, we divide the network to subgraphs with similar \maxhtlc, and use a greedy algorithm to select routes.
The algorithm that we utilized is parametrized by G (a subgraph of the network) and a parameter $\tau_{min}$ that denotes the minimal time (measured in blocks) that we would like paths to be locked for. We assign two types of weights to the edges (used in selecting the routes to attack) to support two different modes of attack. The greedy algorithm selects routes one by one, constructing each by consecutively picking high weighted edges.
In the first mode we seek to freeze as much liquidity as possible and use the channel capacity as the weight. \change{In the second mode, we seek to disconnect as many pairs of nodes and use the unweighted betweenness centrality measurement of edges as the weight (taking inspiration from the Girvan-Newman Algorithm~\cite{girvan2002community}).}
A concise description of the algorithm follows (Appendix~\ref{alg} provides a detailed description):
\begin{enumerate}
    \item Pick a channel of maximal weight (capacity/betweenness).
    \item Extend it to a route by repeatedly choosing an adjacent channel of maximal weight which meet the constraints of having similar \maxhtlc value and maintaining route validity (maximum route length and desired route locktime $\tau_{min}$).
    \item Remove the route channels from the graph.
    \item Repeat until all channels are exhausted.
\end{enumerate}

The result is a partition of G's channels into disjoint routes that can be paralyzed for at least $\tau_{min}$ blocks.
Note that routes produced by the algorithm are circular (from the attacker to itself) and require two attacker channels: to begin and end each route.

For many channels in the network, the value set for \cltvdelta is different depending on the direction we traverse the channel (this is because nodes may have set different values for this parameter). Our greedy approach excelled at picking directions with lower \cltvdelta values naturally, which allows it to form longer routes that paralyze more channels simultaneously. 
Other approaches that we explored, such as iterating over a single channel back and forth to form a long path, resulted in slightly worse performance. 

\change{The greedy algorithm does not optimize over the \htlcmsat values when picking routes (Algorithm~\ref{alg::1}). The values set for this parameter are extremely low and their impact on the total cost is minor.
}

\subsection{Evaluation}

We run the attack locking channels for at least 3 days ($\tau_{min}=432$). We begin by attempting to freeze up a large amount of liquidity (setting the weight in Algorithm~\ref{alg::1} to the channel's capacity). 

We infer for each node, which implementation of the protocol it runs (See Appendix~\ref{tag}), and then partition the network into two sub-graphs:
\begin{enumerate}
	\item  The network graph reduced to \lnd nodes. Which has \maxhtlc defaults that are 483. 
	\item The complementary graph consisting of all channels with at least one \eclair or \clightning node. These use a default \maxhtlc of 30. 
\end{enumerate}

\change{In the implementation inference process we assume most users use the default values for the \maxhtlc parameter. This assumption is reinforced by Section~\ref{Statistics}, which presents distributions of other parameter values displaying high correspondence to their default values.}

\begin{figure*}[t!]
    \centering
    \begin{subfigure}[t]{0.5\textwidth}
        \centering
        \includegraphics[scale=0.447, trim={0.15cm 0cm 0.27cm 0.23cm}, clip]
	{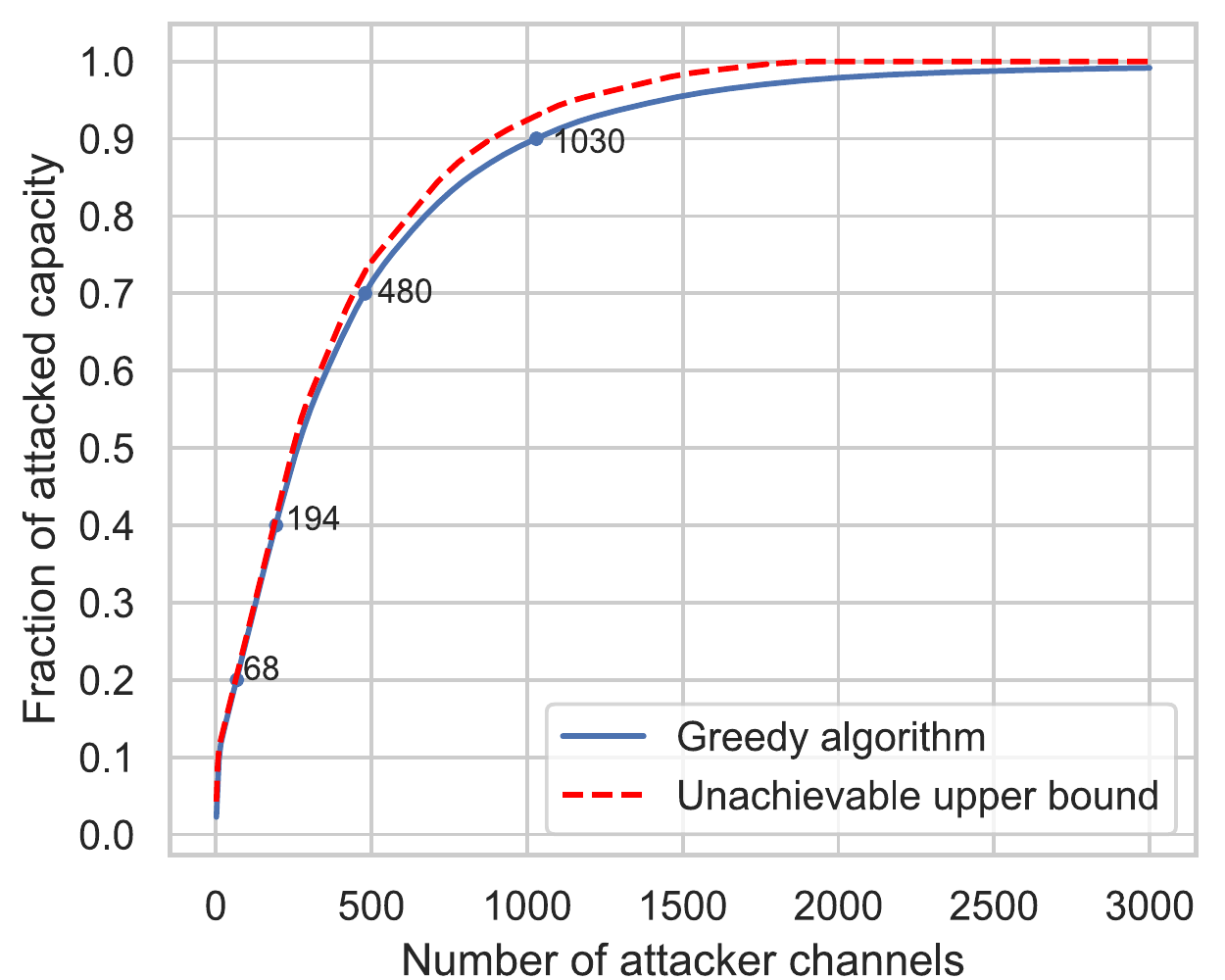}
	\caption{{Fraction of attacked network capacity}}
    \end{subfigure}%
    ~ 
    \begin{subfigure}[t]{0.5\textwidth}
        \centering
        	\includegraphics[scale=0.456, trim={0.22cm 0cm 0.6cm 0.5cm}, clip]{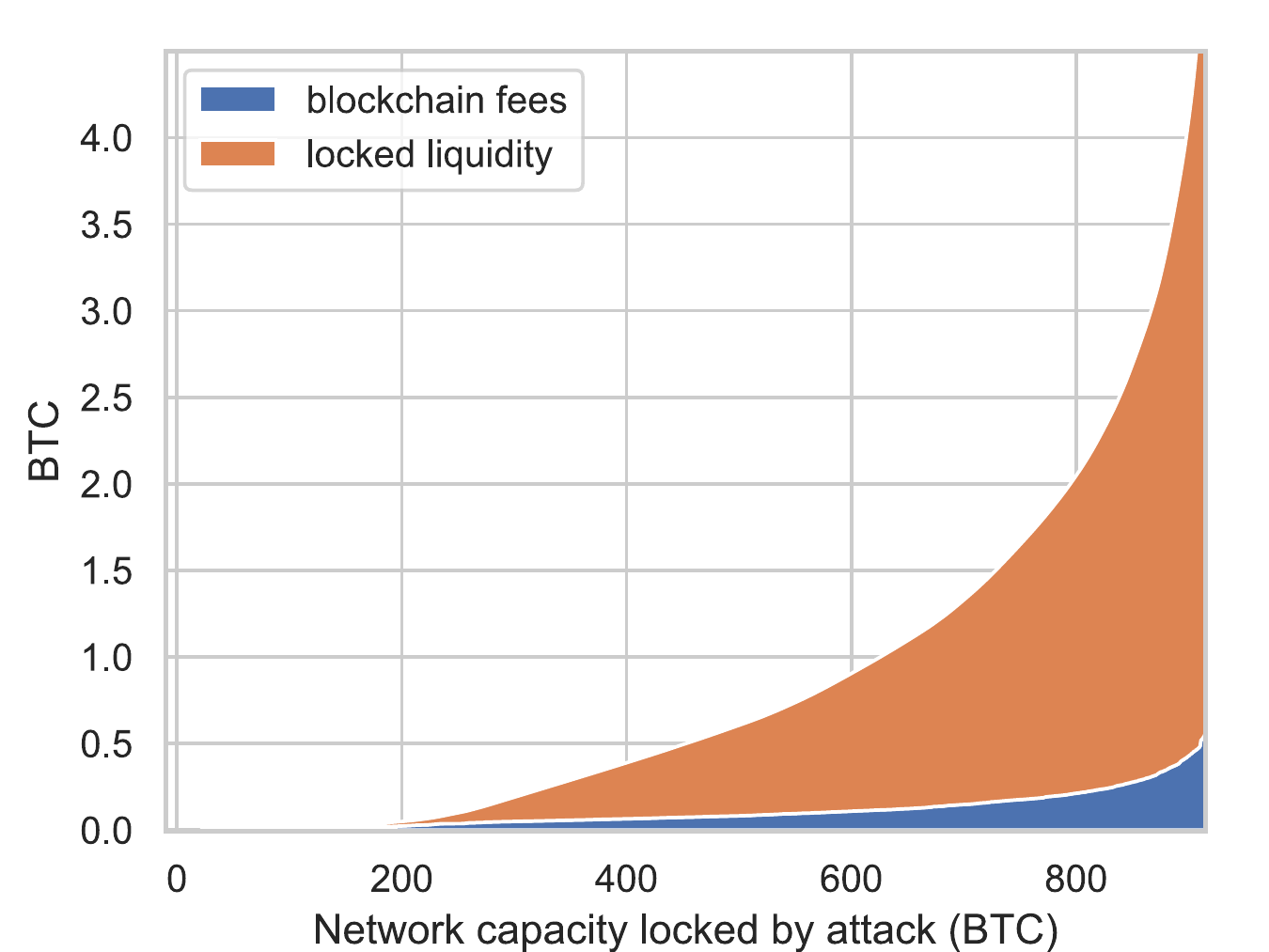}
	\caption{Cost of the attack}
        	
    \end{subfigure}
    \caption{Evaluation and cost of the attack}
     \label{figure:8}
\end{figure*}

We visualize the results in Figure~\ref{figure:8}a, presenting the fraction of the network's capacity that the attacker succeeds in locking as a function of the resources it invests (the number of channels it opened). We find for example that the attacker can lock 20\% of the network's capacity using only 68 channels, and can lock 90\% using 1030 channels. 
We notice that the greedy algorithm is almost optimal on our graph. To do so we compare to an unachievable upper bound which is calculated as follows:
The maximum allowed route length is 20. The attacker uses 2 channels to attack any route, hence it can attack at most 18 channels per route. We sort the channels by their capacities and use the highest capacity edges first, disregarding the constraint that paths are connected correctly.

We estimate the attack's costs, by considering two types of costs:
\begin{enumerate}
	\item The cost of opening channels. The attacker pays the fee required to place channel funding transactions on the blockchain. We estimated the cost of opening a channel to be 2.2 USD (the average transaction fees observed on the date of the snapshot which the evaluation was performed on)~\cite{BitInfoCharts}.  
	\item The cost of provisioning channels with liquidity. Attackers must lock enough liquidity for payments they will later request. Locked funds are not spent and will return to the attacker once it completes the attack. 
\end{enumerate}

Figure~\ref{figure:8}b displays our evaluation of the costs. It clearly separates the two types of costs mentioned above (non-refundable blockchain fees and locked liquidity).
Our results show that the attacker can paralyze most of the liquidity in the Lightning Network for 3 days spending less than half a Bitcoin.
We take into account the dust limit configured in the main Lightning implementations which sets a threshold on the payment size below which \htlcs would not be added by nodes.
The locked liquidity cost is mainly affected by the dust limit, the rest of the parameters (min \htlc, base fee, minimal channel capacity) have less of an impact because of their small values.
The costs we estimate above can be further lowered by opening multiple channels with a single on-chain transaction. Once channels are established, the attack may be repeated again and again with no additional cost to the attacker.

\change{
To be able to block a route there needs to be sufficient balance in each channel to allow for a minimum payment (otherwise nodes along the path will reject the payment request). The required balance relies mainly on the dust limit and the \maxhtlc values configured along the route.
}

\begin{figure}[ht]
	\centering
	\includegraphics[scale=0.32]{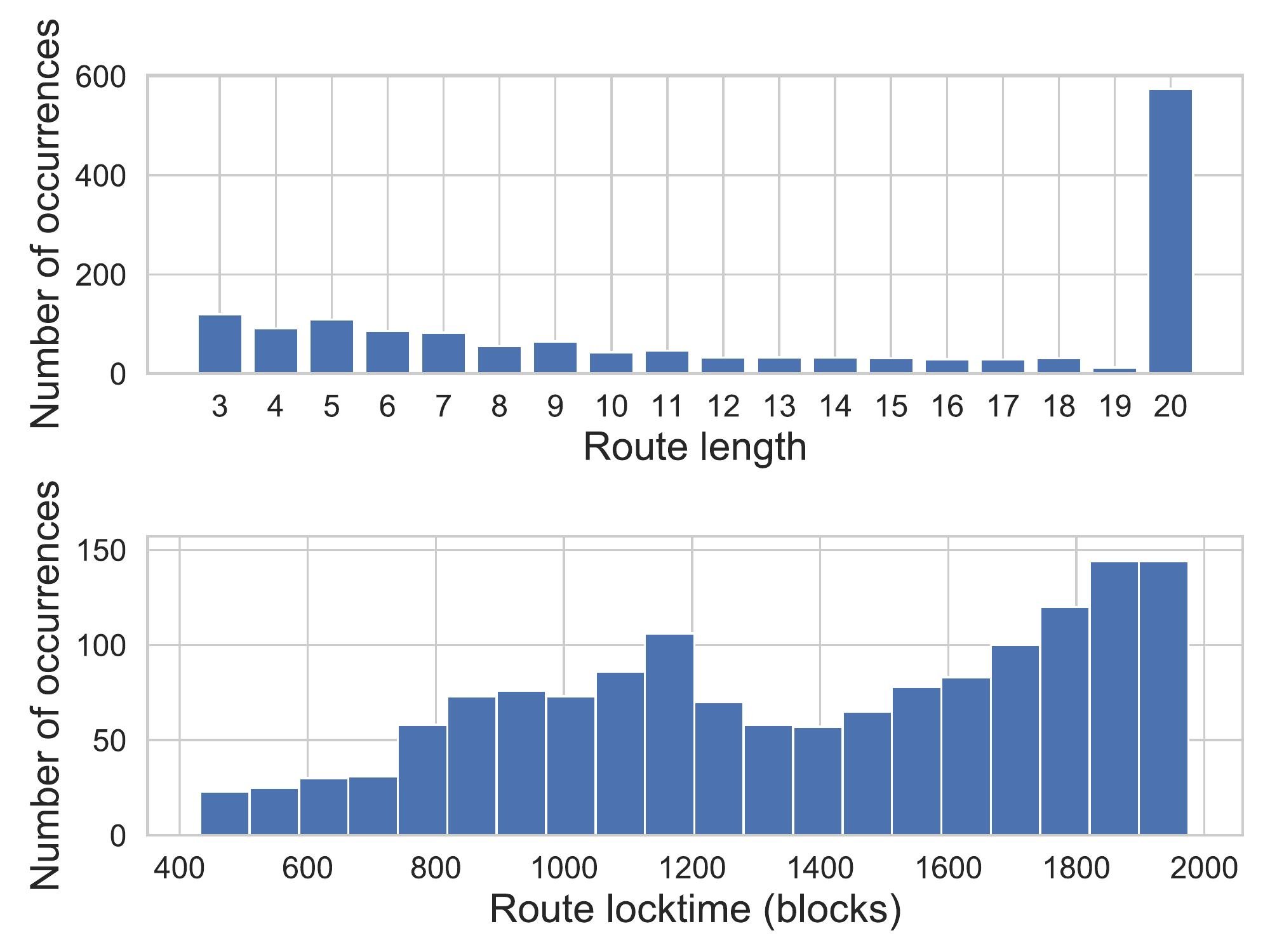}
	\caption{{Histogram of route lengths (including attacker's edges) and route lock times}}
	\label{figure:7}
\end{figure}

We show more details on the attack results in Figure~\ref{figure:7}. The figure shows that the attacker succeeds in attacking long routes (exploiting maximum route length), and that most of the routes are locked for more than the 3 days that were set as the minimal lock time. 

\begin{figure*}[t!]
    \centering
    \begin{subfigure}[t]{0.5\textwidth}
        \centering
       	\includegraphics[scale=0.396, trim={0cm 0.05cm 0.1cm 0.1cm}, clip]
	{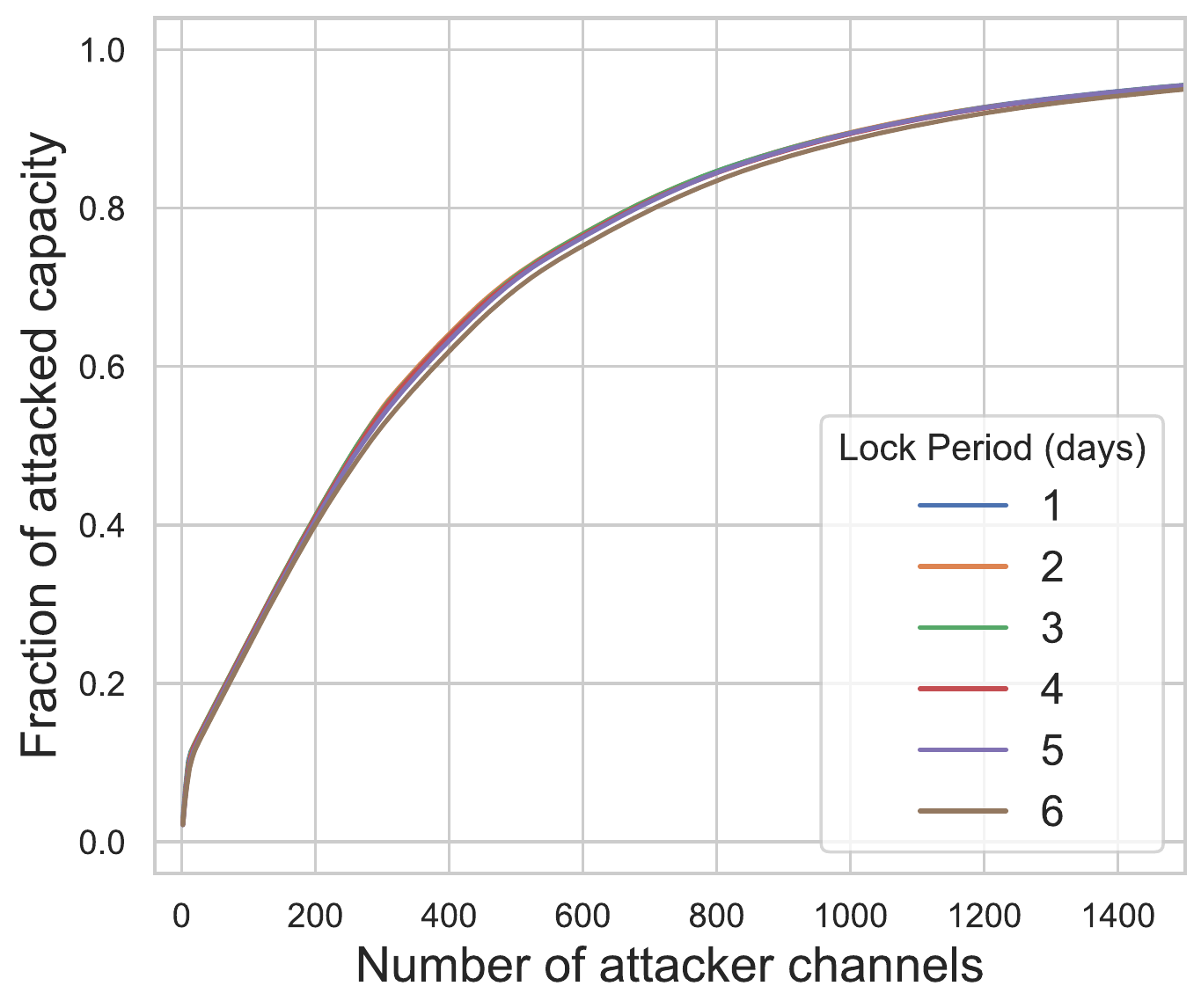}
	\caption{Fraction of attacked network capacity for different lock periods}
    \end{subfigure}%
    ~ 
    \begin{subfigure}[t]{0.5\textwidth}
        \centering
	\includegraphics[scale=0.415, trim={0.2cm 0.1cm 0cm 0.1cm}, clip]
	{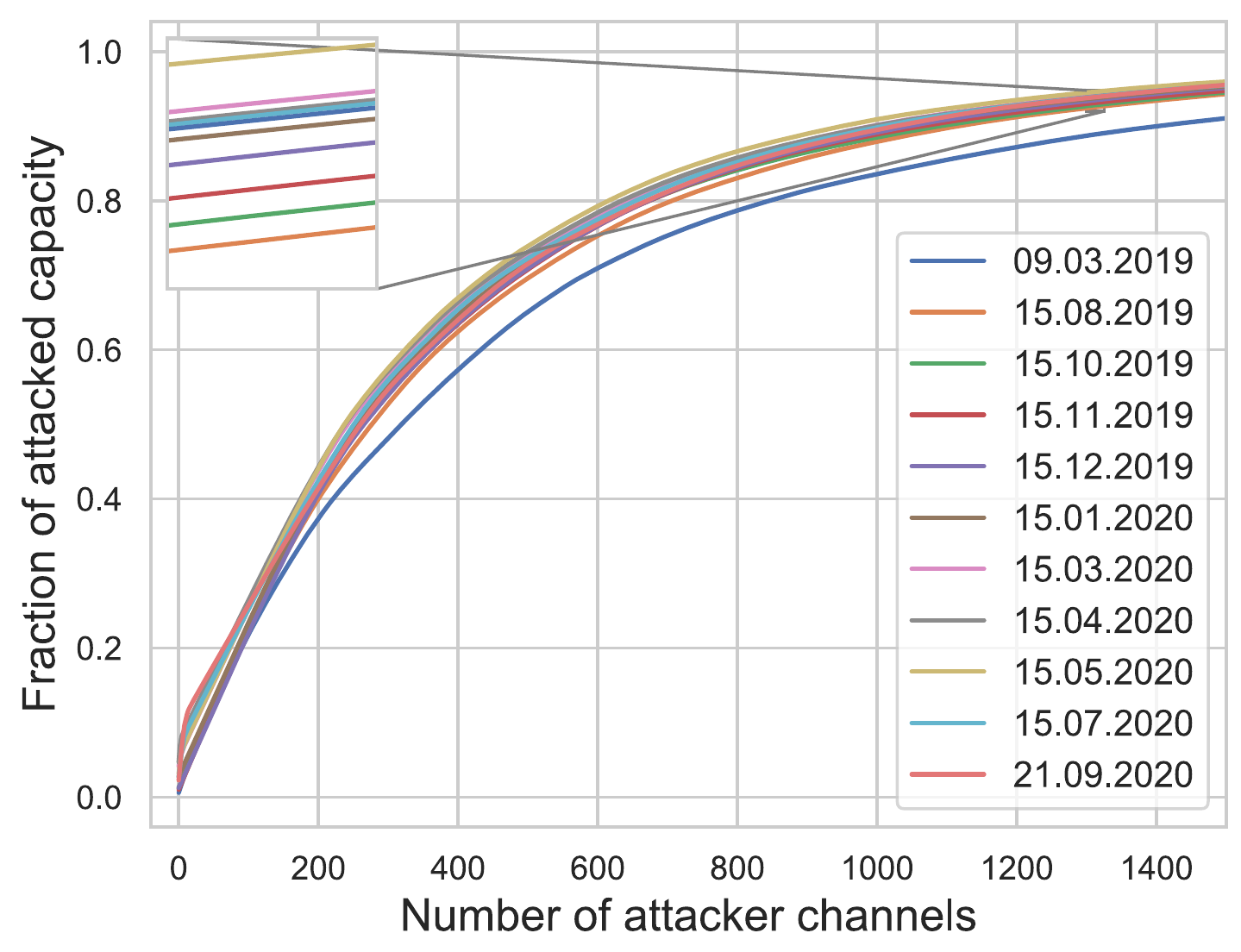}
	\caption{Fraction of attacked network capacity in different snapshots}
    \end{subfigure}
    \caption{Evaluation of the attack}
     \label{figure:10}
\end{figure*}

In Figure~\ref{figure:10}a, we run the attack changing the number of days that channels remain locked for. 
The results indicate that the number of attacker channels required to lock paths for different periods (from 1 to 6 days) differs only slightly. This can be explained by the relation between the large \maxlock (2016 blocks) value, the small \cltvdeltas, and the 20-hop route length constraint. 
In other words, most of the liquidity of the network can be attacked using routes that consist of small \cltvdeltas, allowing the attacker to high timeouts and withhold the payments for a long period.

Figure~\ref{figure:10}b explores how the attack would work on the Lightning Network at different times. We use snapshots taken over several months. The results generally show that the attack gets \emph{easier} as time passes (there is a slight improvement from May 2020). This can be explained by the changes made to default parameters -- increasing \maxlock to 2016 in all implementations and decreasing \cltvdelta from 144 to 40 in \lnd. Both changes make it easier to construct long routes with high timeouts.


\begin{figure}[ht]
	\centering
	\includegraphics[scale=0.52, trim={0.2cm 0.03cm 0.2cm 0cm}, clip]
	{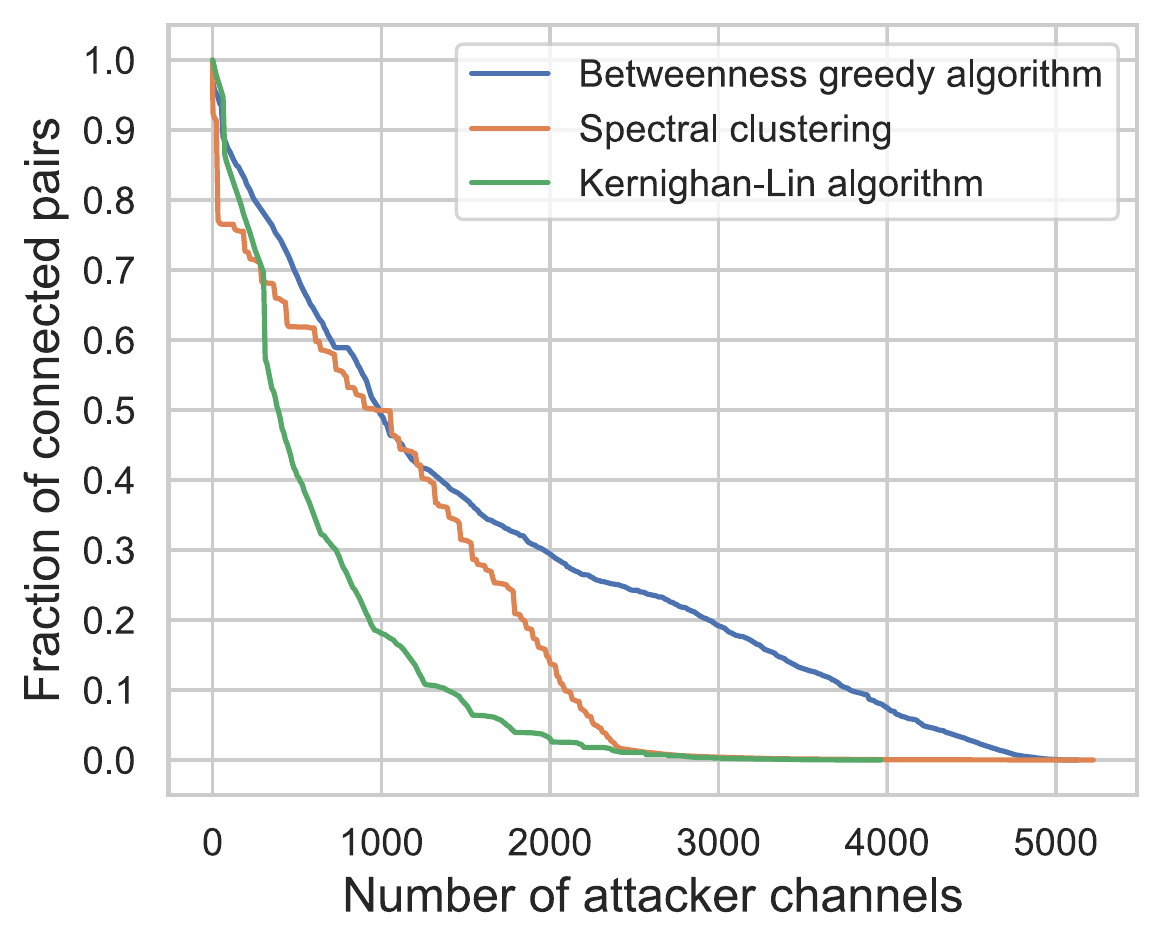}
	\caption{Fraction of connected pairs of nodes in the network}
	\label{figure:9}
\end{figure}

In figure~\ref{figure:9}, we show how the attack affects connectivity between nodes in the network. We explored several algorithms to select the attacked routes and present them in Figure~\ref{figure:9}. The algorithms we explored are: Using Algorithm~\ref{alg::1} -- a greedy algorithm which picks channels with high betweenness centrality. The second approach utilizes spectral clustering to repeatedly cut the large connected component using an eigenvector corresponding to the second smallest eigenvalue (Fiedler vector) of the Laplacian matrix of the largest connected component. The sign of the coordinates partitions the vertices of the graph into 2, defining a cut~\cite{fiedler1989laplacian}. Finally, we used a simplified version of Kernighan-Lin algorithm~\cite{kernighan1970efficient} that starts with an arbitrary partition that separates $1/4$ of the nodes and greedily swaps nodes across the cut to minimize the cut. This yielded the best results.

While before the attack almost all pairs of nodes ($>$97\%) are connected, using only 32 attacker channels we disconnect 23\% of the pairs in the network, while with 385 channels we disconnect 50\% of the pairs.
We stress that right now different Lightning implementations try only a small handful of paths~\cite{tochner2019hijacking}, so even a large fraction of nodes that we noted as connected will not be able to route payments between them.


\section{Attacking Hubs - Attack on a Single Node}\label{attack2}
In this section we consider an attack aimed at disconnecting a single node from the network for an extended period of time.
Here, the adversary connects to the victim and paralyzes its adjacent channels one by one using the following steps:
\begin{enumerate}
    \item The adversary connects to the victim with a new channel.
    \item It then initiates a payment to itself via a route that begins with its connection to the victim, and then traverses a single target channel back and forth multiple times, before returning to the attacker. Surprisingly, such paths that traverse channels back and forth are indeed possible (see Appendix~\ref{proof}). 
    \item The attacker makes multiple payment requests over this path until the target channel reaches \maxhtlc. In this case, the attacker's own channel is usually not maxed out, and can be used to attack again.
\end{enumerate}

We note that the attack is still possible to carry out if the victim does not accept direct connections (but at a somewhat lower efficiency). In this case, we would connect to neighbors of the victim.


Once the target channel is paralyzed, we move to the next one and apply the same method. We will need to open a new channel between the adversary and the target node every time that the former reaches its \maxhtlc. Yet, at each payment we withhold only two \htlcs on the adversary's channel while it is possible to reach up to 18 \htlcs in the target channel at the same time. In other words, in order to attack all of the victim's channels, the adversary needs to open a small number of channels relative to the victim's degree.

\subsection{Evaluation}

We evaluate the attack on prominent nodes in the network. The following table 
summarizes our results: 

\begin{center}
\setlength{\tabcolsep}{10pt}
		\resizebox{8,2cm}{!}{%
		\begin{tabular}{ c  c  c  c }
			\hline
			\textbf{Alias} & \textbf{\% of Network} & \textbf{Node's} & \textbf{Attacker}\\ 
			 & \textbf{Liquidity} & \textbf{Degree} & \textbf{Channels}\\ 
			\hline
			ACINQ           & 10.8\%   & 774       & 151            
			\\ 
			 Bitfinex [lnd1]  &  6.4\% & 169 & 19
			\\
			OpenNode      & 4.2\%   & 648   & 88
			\\	
			 Bitrefill     & 3.8\%   & 229   & 39
			\\
			CoinGate  & 3.1\% & 609 & 68
			\\ 
			LNBIG (25 nodes) & 22.2\% & 3835 & 405
			\\ \hline
		\end{tabular}
 		}
    
\end{center}

The names of nodes were taken from our snapshot data directly. The last entry in the table relates to an attack on LNBIG~\cite{lnbig},
a single entity that controls 25 nodes which are extremely central to the network, holding a significant share of the network's capacity in multiple channels. 
We isolate all 25 nodes, without paralyzing links between the nodes themselves. Paths were set so that all links are paralyzed for at least 3 days in each iteration.

\begin{figure}[ht]
	\centering
	\includegraphics[scale=0.3]{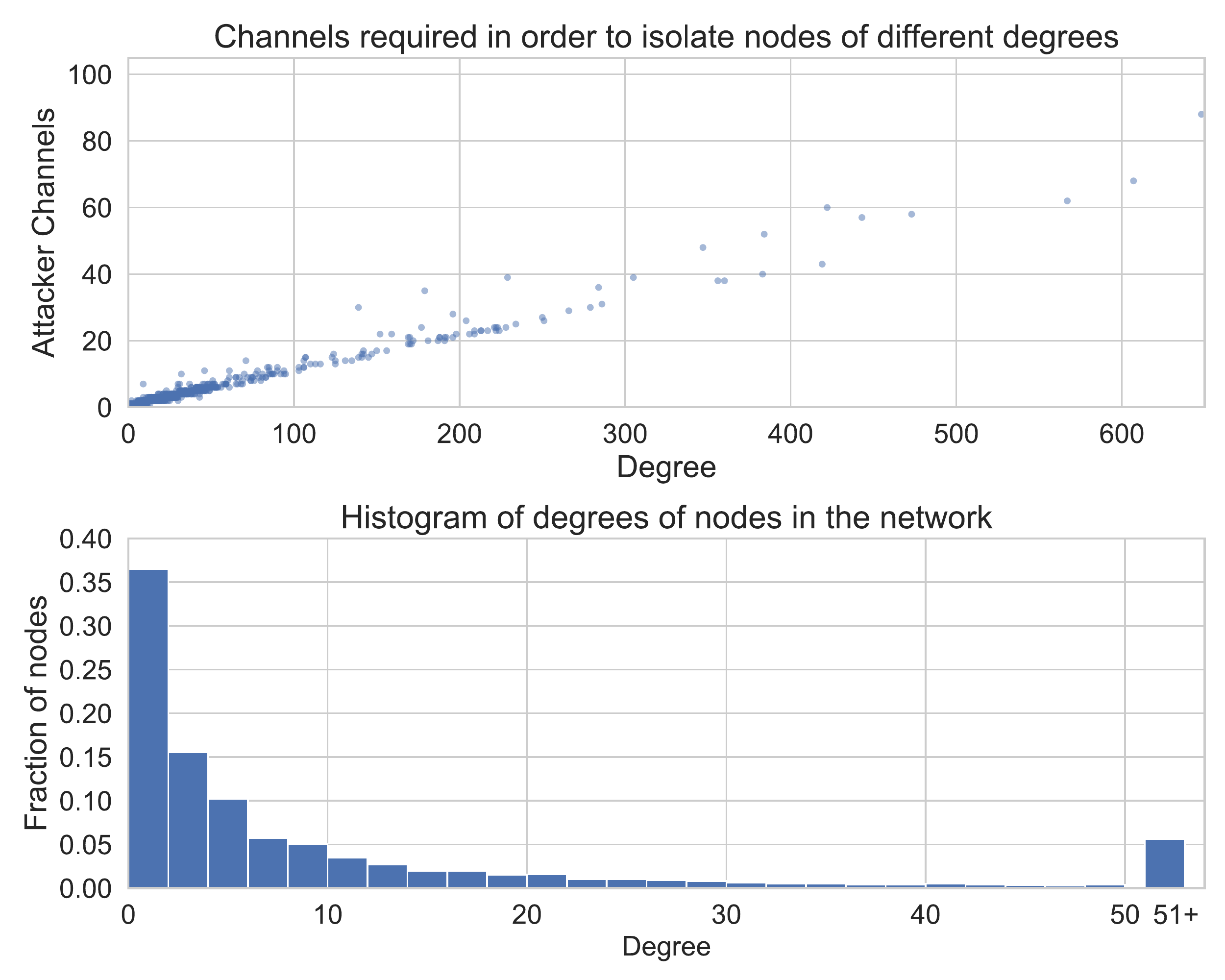}
	\caption{Degree analysis}
	\label{figure:12}
\end{figure}

We evaluated the cost of attack on \emph{all} nodes in the network using a snapshot from September 21st, 2020, isolating each node for 3 days.
Figure \ref{figure:12} presents a histogram of the degree of nodes and shows the relation between the degree and the number of channels attackers needed to perform the attack on each node. Each node is represented by a point in the graph. The number of channels is not directly determined by the degree, because different nodes have set up different values of \cltvdelta.
We see that most nodes have a very low degree and are extremely easy to isolate. Even nodes with high degree, require far fewer channels than the degree to attack. 

In an additional evaluation which is described in Appendix~\ref{degree}, we show that of the 3 main implementations of the Lightning Network, \lnd (the most common implementation) nodes are the easiest to attack.

\section{Mitigation Techniques}\label{Solutions}

\change{The attack and vulnerabilities described in our work continue to be relevant and have been discussed by the Lightning community~\cite{Spamming,pull_request}.} In this section we discuss several proposed adjustments to payment channel network protocols that may help mitigate the attack. Specifically, we discuss some ideas that were raised in the Lightning-dev mailing list~\cite{EmelyanenkoK,Rusty}, as well as our own suggestions. We discuss weaknesses and strengths of each such suggestion.

\paragraph{Enforcing fast HTLC resolution} - This is our most drastic suggestion: While \htlc expiration times allow nodes to remain secure and provide sufficient time to publish transactions to the network, we propose the addition of another timeout mechanism. Specifically, if \htlc secrets are not propagated fast enough from one's neighbor the channel with this neighbor should be closed. 

Each node should announce to its successor in the path its own deadline for resolving the \htlc. The node would then be able to communicate an earlier deadline for \htlc resolution to its next hop. If the timeout arrives, and the \htlc was not fulfilled or canceled, the node will wait for the \htlc to naturally expire but will close the channel with its neighbor. 

To avoid having all channels along the path closed due to a failure to complete the \htlc in time, and specifically to avoid closing channels between compliant nodes, the last node in the path will provide proof of the channel closure to its predecessors (this can be done using a zero-knowledge proof for example).

We stress that this proposed mechanism does not replace the \htlc timeouts that still ensure the safety with regards to the \emph{current} payment. Our mechanism is a way to disconnect misbehaving peers from the network in order to prevent them from repeating the attack many times at no cost. 
We note that it is risky to add behavior that automatically closes channels, and so this proposal warrants further evaluation. We leave this to future work. 

\begin{figure}[ht]
	\centering
    \includegraphics[scale=0.43, trim={0cm 0cm 0cm 0cm}, clip]{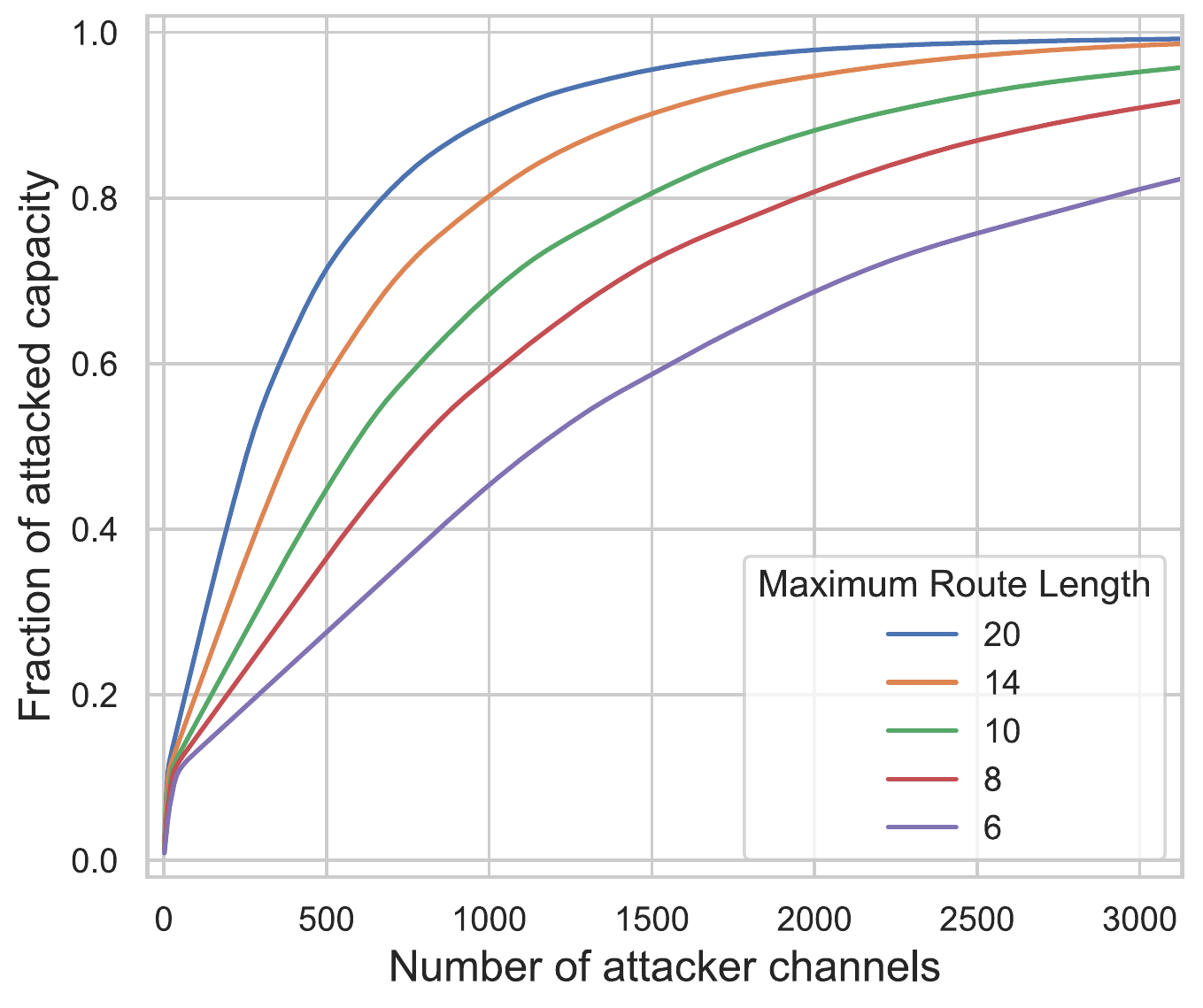}
	\caption{Fraction of capacity attacked for different max route lengths}
	\label{figure:14}
\end{figure}

\paragraph{Reducing route length} - We suggest lowering the maximum allowed route length (currently 20 hops)\change{, as suggested in previous work~\cite{perez2019lockdown}}. The network graph is a small world  network~\cite{rohrer2019discharged} - it is highly connected, and a smaller number of hops should still suffice. We point out that shortest paths between nodes in the network have an average of less than 3 hops and that the network diameter is $\sim 6$~\cite{rohrer2019discharged,seres2019topological}, which are significantly lower than the 20 allowed hops. In Figure~\ref{figure:14}, we show the fraction of successfully attacked capacity (with respect to the attack described in Section~\ref{attack1}), assuming that different max route lengths are allowed. The figure shows that attackers need many more channels to attack if they are forced to use shorter route lengths.

\paragraph{Setting number of max concurrent payments based on trust level} - Currently, each node configures \maxhtlc to bound the maximum transfers it is willing to hold concurrently. 
Most nodes use the default value configured by the implementation they run. In all cases this value should not exceed the number 483 which is derived from the blockchain's limitations.
We suggest changing the way nodes configure this parameter, adjusting the value according to the level of trust they have in particular peers. Setting a high \maxhtlc for some peer effectively allows it to route many concurrent payments through your node and to do more damage if it is malicious. Therefore, newly created channels with unknown and untrusted nodes should default to a low \maxhtlc. 
\change{A new pull request has been opened recently (Aug 2020) promoting the basis of our proposal, allowing nodes to limit their exposure to the maximum number of concurrent \htlcs~\cite{pull_request}.}

\paragraph{Loop Avoidance} - As our experiments show (Appendix~\ref{proof}), it is possible to construct paths that visit the same node several times, including traversals of the same channel back and forth. It is relatively simple for nodes to disallow such paths. Since \htlcs that belong to the same path use the same hash, they can be easily recognized and rejected. This will make our specific technique to isolate individual nodes harder to carry out, but may not solve the issue entirely.


\section{Related Work} \label{related}

A DDoS attack on the Lightning Network occurred in March of 2018. Many nodes were flooded with traffic and around 200 Lightning nodes were taken offline~\cite{DDoS}.
Several studies explore sophisticated attacks on the Lightning Network. Some focus on privacy issues~\cite{herrera2019difficulty,tang2019privacy}, and others on isolating nodes~\cite{rohrer2019discharged,perez2019lockdown} or disrupting the network in other ways. 
\change{Rohrer et al.~\cite{rohrer2019discharged} explored} an attack that disrupts the liquidity balance of channels. The attacker initiates payments that move all the liquidity to one side, effectively blocking payments in that direction (payments in the other direction are still possible). 
Our attack differs from this attack ~\cite{rohrer2019discharged}, as they require direct connections to the victim node, as well as locked liquidity in high amounts (up to the liquidity the victim has), in addition to the payment of fees for large transactions. 

A similar attack uses payment griefing but avoids paying the fees~\cite{perez2019lockdown}. In this variant of the attack, the attacker still sends a payment in one direction that unbalances the channel in order to isolate a node. This time it withholds the \htlc pre-image in order to lock the amount, and never really executes the payment. Unlike our attack, this still requires amounts of locked funds that match the amounts being locked in the victim channel, but does indeed avoid paying most of the fees (channel establishment is still needed). 

\change{Tochner et al.}~\cite{tochner2019hijacking} presents a denial-of-service attack based on route hijacking within the Lightning Network. 
They show how connecting with few channels to the network offering low fees draws most of the routes which yields a potent attack. Our work does not rely on the routing strategy of nodes to attack the network. 

\change{
In a work parallel to ours, Tikhomirov et al.~\cite{tikhomirov2020quantitative} quantify the effect of several attacks on the Lightning Network. They additionally discuss the limitation on the number of concurrent \htlcs and describe the attack vector and its effect on the network's scalability. We describe how to leverage the attack and provide deeper analysis.  
}

The privacy of payments in the Lightning Network is known to be relatively weak. Discovering the current liquidity balance of a channel can be accomplished using techniques from \change{Herrera-Joancomarti et al. research}~\cite{herrera2019difficulty}. 
\change{Tang et al.}~\cite{tang2019privacy} explore the tradeoff between privacy and utility in PCNs, considering adding noise to channels as well (which adds privacy, but lowers efficiency).

The structural properties of the Lightning Network and its topology in the context of the network's robustness have been studied in \change{several studies}~\cite{seres2019topological,lee2020robustness}. 

A technique to lower the delay of \htlc expiration is described by \change{Miller et al.}~\cite{miller2019sprites} and would make our attack less severe in this context. 
This sort of technique is not applicable to the Bitcoin blockchain, due to its more limited scripting language. 


Several protocols improving upon privacy issues in off-chain payment channels are suggested in \change{previous studies}~\cite{green2017bolt,heilman2017tumblebit,malavolta2017concurrency}, \change{as well as} advances in payment channel networks like the addition of watchtowers~\cite{mccorry2019pisa,avarikioti2018towards}.

\change{McCorry et al.}~\cite{mccorry2016towards} presents a technical overview of Bitcoin's payment channel networks. 
Additional work on off-chain protocols can be found in  \change{SoK survey}~\cite{gudgeon2019sok}.

\section{Conclusions and Future Work} \label{conclusions}
In this paper we discussed a fundamental vulnerability that arises in payment channel networks as part of the construction of trust-less multi-hop payments.
We presented three types of attacks: the first aims to lock as many high liquidity channels as possible for an extended period, the second disconnects as many pairs of nodes as possible in the network, and the third isolates hubs from the rest of the network. We evaluated these attacks over the Lightning Network.
We examined the network's properties and different parameters set by the three main implementations of the Lightning Network. We showed how recent changes in default parameters agreed upon by Lightning Devs have made the attack easier to carry out.
Our results show that it is possible to disrupt the Lightning Network at a relatively low cost.

Further work must be conducted in order to mitigate this type of attack. We suggested several solutions to reduce the success rate of these attacks, but such mitigation is generally harder due to the nature of the attack: it relies on several fundamental properties of payment channel networks, and the blockchain.


\section{Acknowledgments}
      We thank Itay Cohen, Nir Lavee and Zvi Yishai for providing improvements in our  network partitioning algorithms and analysis.
    
	This research was supported by the Israel Science Foundation (grant 1504/17) and by a grant from the HUJI Cyber Security Research Center in conjunction with the Israel National Cyber Bureau.

\bibliographystyle{splncs04}
\bibliography{main}

\appendix
\clearpage
\addcontentsline{toc}{section}{Appendices}
\renewcommand{\thesubsection}{\Alph{subsection}}

\subsection{Tagging Nodes by Implementation}\label{tag}
Our attack is based on overloading channels to their maximum \htlc capacity, which is determined by the minimum \maxhtlc of the peers in the channel. 
The \maxhtlc is a configured parameter set by the node's owner. The default \maxhtlc\ in the mainnet configuration for different implementations is 483 for \lnd nodes, and 30 for \clightning and \eclair nodes.

The information of which client a node runs and what value of $\mathtt{max\_concurre}$-$\mathtt{nt\_htlcs}$ it uses is not accessible publicly.
Hence, we use the data nodes do publish via $\mathtt{channel\_update}$ messages in order to infer which implementation they run and deduce the \maxhtlc defaults. Here, we rely strongly on the assumption that most users do not change default values too much. This assumption is supported by the statistics we show in Section~\ref{Statistics}. 

We perform classification using defaults from the table in Section~\ref{defaults} in order to infer the implementation of each node. Where nodes deviated from defaults, we use a score that is weighted according to the parameter that was changed in order to infer the most likely original implementation. 
The weights we have used are:
\begin{itemize}
    \item \cltvdelta~: 0.75
    \item \htlcmsat~: 0.2
    \item \feeproportional~: 0.05
\end{itemize}

For each node in the network, we check the compatibility of these parameters in the policies of its channels with the implementation defaults (using the weights) and decide the label.
\footnote{The results were robust to the use of different classification methods that we tried. We found that $46\%$ of nodes did not change \emph{any} of their channels default values and that $57\%$ of all \emph{channels} were created with default values by at least one of their peers.}

\begin{figure}[ht]
	\centering
	\includegraphics[scale=0.38 , trim={0cm 0cm 0cm 0cm}, clip]{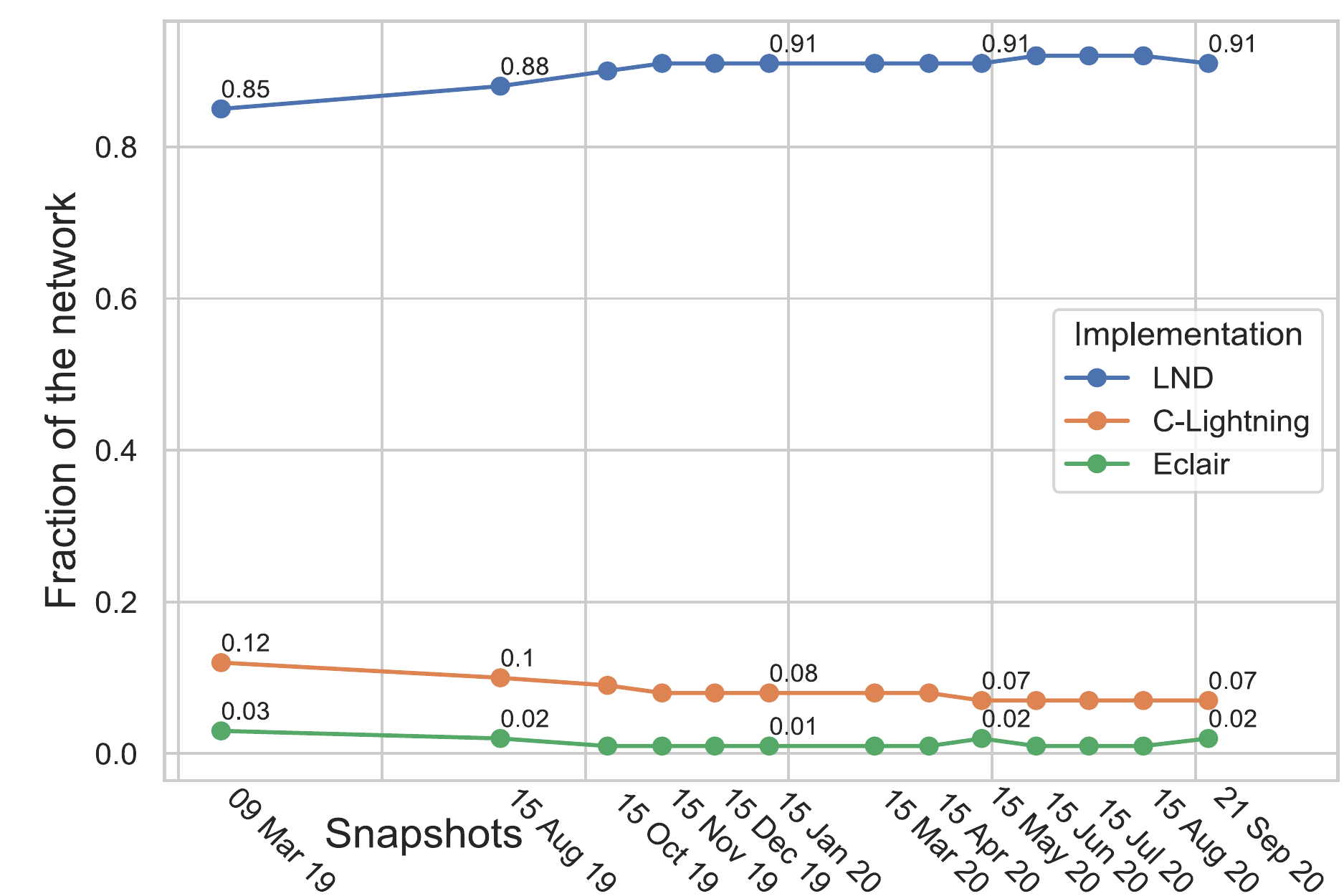}
	\caption{Deduced implementation distribution}
	\label{figure:4}
\end{figure}

The results of the implementation inference at different times are presented in Figure~\ref{figure:4}.
We see that our analysis resulted in tagging approximately 91.3\% of the nodes as \lnd.
When we then reduce the network graph to nodes labeled \lnd alone, we find that we remain with 73\% of the network's original capacity (in BTC). Hence, \lnd nodes are both the most common nodes, and also the ones that hold most of the liquidity in the network. 

\subsection{Proof of Concept Implementation} \label{proof}
To demonstrate the viability of the attacks, we conducted several experiments using \clightning nodes that were running over a separate Bitcoin network (in regtest mode). We detail the main experiments conducted---additional ones appear in code. The source code of our implementation including all experiments is available at \href{https://github.com/ayeletmz/Lightning-Network-Congestion-Attacks}{github}\footnote{\href{https://github.com/ayeletmz/Lightning-Network-Congestion-Attacks}{https://github.com/ayeletmz/Lightning-Network-Congestion-Attacks}}.

We begin with an experiment that shows that we can maintain a single \htlc ``live'' for an extended period, and that we can then revert the state without having any of our channels closed. 
\begin{experiment}[Maintaining a Single \htlc]\label{exp1}
	We set up a simple network Attacker1 $\leftrightarrow$ Alice $\leftrightarrow$ Bob $\leftrightarrow$ Attacker2.  Then Attacker1 initiates a single payment to Attacker2. Attacker2 does not respond with the \htlc secret immediately, but instead waits. We consider two different waiting periods: 
	\begin{enumerate}
		\item Waiting past the \htlc timeout. In this case Bob closes the channel with Attacker2.
		\item Waiting one block before the \htlc timeout, and sending an $\mathtt{update\_fail\_htlc}$ message. In this case the payment is canceled, the corresponding \htlcs along the entire path fail, but all channels remain open. 
	\end{enumerate} 
\end{experiment}

The following experiment that we conducted demonstrates our ability to block payments over a channel between victims Alice and Bob.
\begin{experiment}[Blocking the Victim's Channel]
	We repeat the setup of Experiment\ref{exp1}: Attacker1 $\leftrightarrow$ Alice $\leftrightarrow$ Bob $\leftrightarrow$ Attacker2. 
	We create 483 different payments from Attacker1 to Attacker2. Again, Attacker2 does not instantly respond with the secret. 
	
	We now try to establish one more additional payment from Alice to Bob. The payment fails. Just before \htlc expiration, Attacker2 responds with $\mathtt{update\_fail}$-$\mathtt{\_htlc}$ messages for all payments and now, an additional payment does succeed. 
\end{experiment}
The above experiment also blocks the payment from Alice to Bob if some of the 483 payments are in the reverse direction (from Attacker2 to Attacker1), demonstrating that the \maxhtlc limit applies to \htlcs in either direction.

We additionally tried the experiment above in paths that contained loops (even including back and forth traversals of a single channel). These are all allowed by nodes. We used this in a proof of concept experiment to attack a single hub.
\begin{experiment}[Back and forth attack on a single hub]\label{exp3}
	We set up the network Attacker1 $\leftrightarrow$ Hub $\leftrightarrow$ Node $\leftrightarrow$ Attacker2. Attacker1 initiates payments to itself in the following route, and does not respond with the secrets: 
	\begin{enumerate}
		\item  Attacker1 $\rightarrow$ Hub.
		\item 9 times back and forth on Hub $\leftrightarrow$ Node.
		\item Hub $\rightarrow$ Attacker1.
	\end{enumerate}
	After 26 such payments, Hub $\leftrightarrow$ Node holds $26\cdot18=468$ unresolved payments. Sending an additional payment will fail ($27\cdot18=486 > 483$). Hence, Attacker1 sends 2 additional payments that are meant to fill up the remaining \htlc quota:
	\begin{itemize}
		\item Similarly to the previous paths, only going back and forth 7 times on the channel connecting Hub $\leftrightarrow$ Node
		\item  Finally, a single payment to Attacker2: Attacker1 $\rightarrow$ Hub $\rightarrow$ Node $\rightarrow$ Attacker2 
	\end{itemize}
	
	We now try to establish a payment from Hub to Node, which fails, confirming that we did paralyze this channel. A payment from Attacker1 to Hub succeeds. In fact, Attacker1 succeeds sending 428 ($483-27\cdot2-1$) more payments to Hub while Hub $\leftrightarrow$ Node is blocked. These ``free'' 428 payments may be used to attack other channels connected to this Hub.
\end{experiment}

In the next experiment we tried paths with varying \maxhtlc values, and verified that the minimal value constrains such paths. We further checked that only the edge with the minimal value is fully locked.
\begin{experiment}[Varying \maxhtlc]
	We set up the network Attacker1 $\leftrightarrow$ Alice $\leftrightarrow$ Bob $\leftrightarrow$ Carol $\leftrightarrow$ Attacker2. Attacker2 does not respond with the \htlc secrets. Attacker1, Alice, Carol and Attacker2 have $\mathtt{max\_concurrent\_ht}$-$\mathtt{lcs}$ configured as 483, while Bob configured \maxhtlc to be 30.
	We create 30 different payments from Attacker1 to Attacker2.
	An additional payment from Attacker1 to Attacker2 fails.
	An additional payment including Bob in the route, fails.
	An additional payment from Attacker1 to Alice, or from Attacker2 to Carol succeeds.	453 additional payments from Carol to Attacker2 are accepted and wait for Attacker2 to respond.
\end{experiment}

In the additional experiments, we also verified that paths are indeed limited to 20 hops and to 2016 block lock-time in total (aggregated over the entire path).

\subsection{Greedy Algorithm Attack} \label{alg}

We describe in Algorithm~\ref{alg::1} the greedy algorithm that we utilize.
The algorithm receives as input:
\begin{itemize}
    \item \textbf{G} - a subgraph of the network.
    \item \textbf{$\tau_{min}$} - the minimal time in blocks that we would like paths to be locked for.
\end{itemize}
and outputs $\mathcal{R}_G$ --- a partition of G's channels into disjoint routes that can be paralyzed for at least $\tau_{min}$ blocks.

\begin{algorithm}
    \DontPrintSemicolon
    \SetKwProg{Fn}{Function}{:}{}
    
    \SetKwFunction{FChooseRoutes}{ChooseRoutes}
    \Fn{\FChooseRoutes{$G$, $\tau_{min}$}}{
        $\mathcal{R}_G \gets \emptyset$ \textcolor{gray}{\Comment*[l]{disjoint routes list}}
        $\mathcal{C}_G \gets E(G)~channels$ \;
         \While{$\mathcal{C}_G \neq \emptyset$}{
             
              $(n_{start}, n_{next}) = \mathcal{C}_G .pop\_max\_weight()$ \;

            \If{$cltv\_expiry\_delta(n_{next}) < cltv\_expiry\_delta(n_{start})$} {
                         $swap(n_{next},n_{start})$ \;
                    }              
            $route \gets (n_{start}, n_{next})$ \;
              \Do{$\mathcal{L} \neq \emptyset~and~ route.length < 19$}{
                $n_{cur} = n_{next} $ \;

                $\mathcal{N} = Neighbors(n_{cur}, \mathcal{C}_G)$ \;
                $\mathcal{L} = \{(n_{cur}, n) | n \in \mathcal{N} , CanExtendRoute(n, route, \tau_{min})\}$\;
                \If{$\mathcal{L} \neq \emptyset$} {
                          $(n_{cur}, n_{next}) =  \mathcal{L}.pop\_max\_weight()$\;
                         $route \gets n_{next}$ \;
                         $\mathcal{C}_G.remove((n_{cur}, n_{next}))$
                    }
              }
              $\mathcal{R}_G \gets route$ \;
         }
          \Return $\mathcal{R}_G$ \;
     }\;
         
         \textcolor{gray}{\Comment*[l]{checks if adding the node keeps the route locked for at least $\tau_{min}$ blocks}}
      \SetKwFunction{FCanExtendRoute}{CanExtendRoute}
        \Fn{\FCanExtendRoute{$n$, $route$, $\tau_{min}$}}{
            $node\_delta = cltv\_expiry\_delta(n)$ \;
            $route\_timeout = \displaystyle\smashoperator{\sum_{n_i  \in~\substack{route}}}{cltv\_expiry\_delta(n_i)}$ \;
            \If{$node\_delta \leq 2016 - \tau_{min} - route\_timeout$} {
                         \Return True \;
                    }
             \Return False \;
         }\;
     \caption{Splits G into disjoint routes that can be locked for at-least $\tau_{min}$ blocks}

     \label{alg::1}
\end{algorithm}

 The algorithm utilizes a ``weight'' function that is used to select the channels that are attacked. In some of our experiments we set this to be channel capacities (seeking to freeze as much liquidity as possible) and in some we set the weight function to be the \change{unweighted} betweenness centrality of the edges (seeking to disconnect as many pairs of nodes as we can). In the latter, we take inspiration from the Girvan-Newman Algorithm~\cite{girvan2002community}.

The {\bf ChooseRoutes} method splits G into disjoint routes that can be locked for at least $\tau_{min}$ blocks. Each route is constructed by extracting a high weight channel, picking the direction which requires smaller \cltvdelta for forwarding payments, and completing it to a circular route by repeatedly adding high weight channels (where channels had equal weight, we preferred ones with low \cltvdelta). Only channels that keep the route locked for at least $\tau_{min}$ blocks are considered.

The {\bf CanExtendRoute} method is used to check if an adjacent channel is suitable for extending the route by checking if with it, the route can still be locked for at least $\tau_{min}$ blocks.

We apply Algorithm~\ref{alg::1} separately to subgraphs with similar $\mathtt{max\_concurrent\_}$-$\mathtt{htlcs}$. We unify the outputs \mbox{$\mathcal{R}=\bigcupdot\limits_{G}{\mathcal{R}_G}$} and sort the set by routes weights in decreasing order as we want to attack routes with higher weight first.

\subsection{Additional Node Isolation Evaluation} \label{degree}
In Section~\ref{attack2} we evaluate the attack of disconnecting a single node from the network for an extended period of time.

\begin{figure}[ht]
	\centering
		\includegraphics[scale=0.5, trim={0.1cm 0cm 0cm 0cm}, clip]{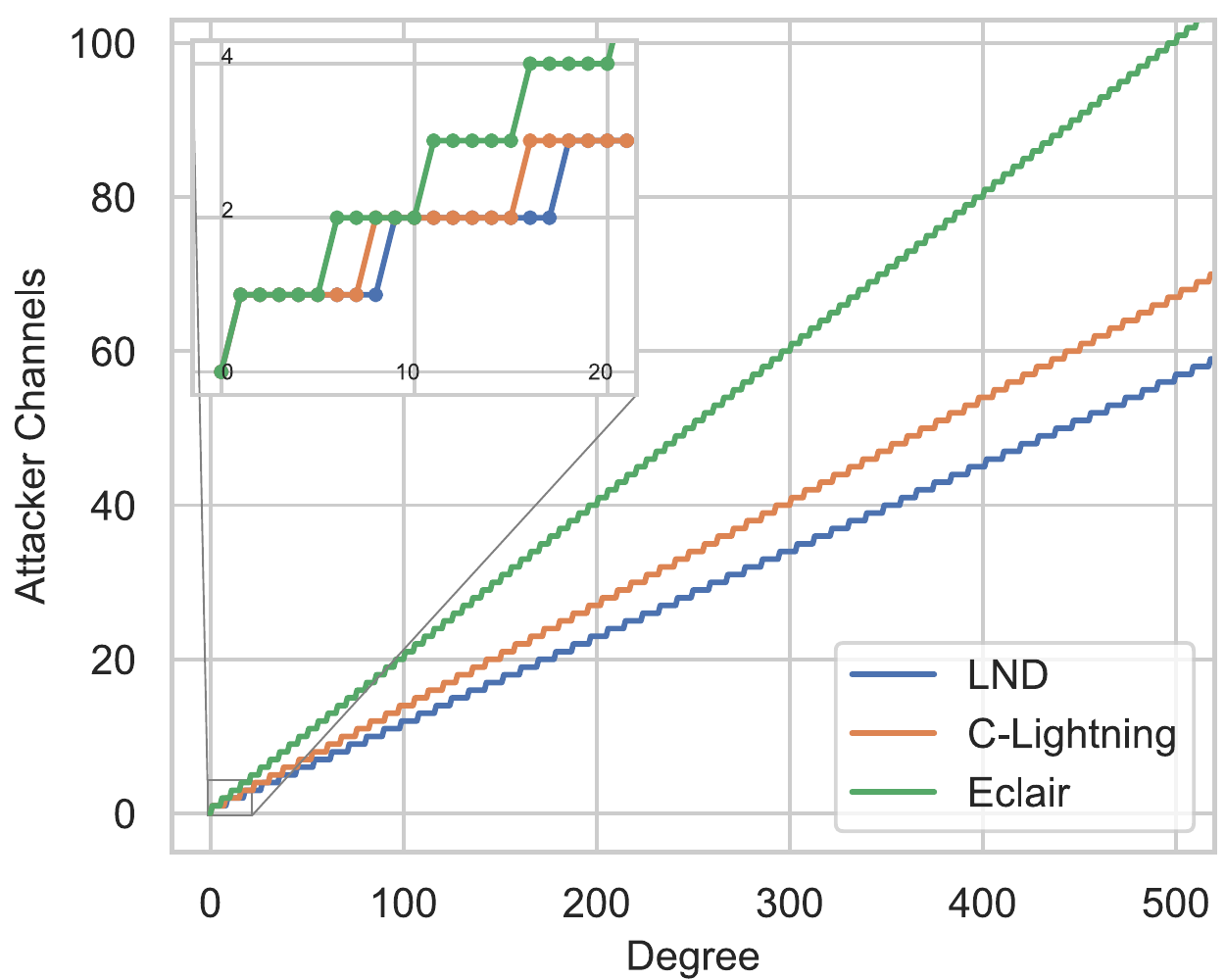}
	\caption{Implementation analysis - number of channels required in order to isolate nodes of different degrees}
	\label{figure:13}
\end{figure}

We present an additional evaluation, estimating the cost of isolating nodes running one of the major implementations, assuming default values are used by it and its neighbors. We calculate the number of channels the attacker needs to open in order to isolate a node for 3 days for different degrees. We present the results in Figure~\ref{figure:13}. Notice that implementations differ due to their different default values. We recall from Appendix~\ref{tag} that $\sim91.3\%$ of nodes run \lnd. Figure~\ref{figure:13} shows that these are the easiest to attack. 

\eclair is hardest to attack due to its higher default \cltvdelta value.  \lnd's value of \cltvdelta is higher than \clightning (40 vs 14), but it is still easier to attack due to its different \maxlock value (\cltv values were low enough so that they did not form a constraint---the number of hops was the main restriction on path length).

\end{document}